\documentclass[longnamesfirst,apjl]{emulateapj}
\usepackage{apjfonts,natbib}
\tighten

\received{2003 July 20}
\accepted{2003 September 26}

\slugcomment{The Astrophysical Journal Letters, in press}
\shortauthors{BAUER \& BRANDT}
\shorttitle{CHANDRA AND HST OBSERVATIONS OF IC~10 X-1}

\begin{document}

\title{Chandra and HST Confirmation of the Luminous and Variable X-ray Source
  IC 10 \hbox{X-1} as a Possible Wolf-Rayet, Black-Hole Binary}

\author{Franz~E.~Bauer\altaffilmark{1} and
W.~N.~Brandt\altaffilmark{1} }

\altaffiltext{1}{Department of Astronomy \& Astrophysics, 525 Davey Lab, 
The Pennsylvania State University, University Park, PA 16802.}

\begin{abstract}
We present a {\it Chandra} and {\it HST} study of IC~10 X-1, the most
luminous X-ray binary in the closest starburst galaxy to the Milky
Way. Our new hard X-ray observation of \hbox{X-1} confirms that it has
an average 0.5--10~keV luminosity of $1.5\times10^{38}$ erg~s$^{-1}$,
is strongly variable (a factor of $\approx$2 in $\la$3~ks), and is
spatially coincident (within 0\farcs23$\pm$0\farcs30) with the
Wolf-Rayet (WR) star [MAC92]~17A in IC~10.  The spectrum of X-1 is
best fit by a power law with $\Gamma\approx1.8$ and a thermal plasma
with $kT\approx1.5$~keV, although systematic residuals hint at further
complexity. Taken together, these facts suggest that X-1 may be a
black hole belonging to the rare class of WR binaries; it is
comparable in many ways to Cyg~X-3. The {\it Chandra} observation also
finds evidence for extended X-ray emission co-spatial with the large
non-thermal radio superbubble surrounding X-1.
\end{abstract}


\section{Introduction}\label{introduction}

IC~10 is a metal-poor \citep[$Z\approx0.15Z_{\odot}$;][]{Lequeux1979},
barred dwarf irregular in the Local Group. Although hampered by
uncertain reddening corrections, reliable optical and infrared
distance estimates place IC~10 at 0.6--0.8~Mpc \citetext{e.g.,
\citealp{Saha1996}; \citealp{Sakai1999}; \citealp{Borissova2000}; we
adopt 0.7~Mpc}, indicating that it is the nearest starburst galaxy
to the Milky Way. IC~10 is notable for its vigorous star formation
\citetext{$\approx$0.03~$M_{\odot}$ yr$^{-1}$ kpc$^{-2}$; e.g.,
\citealp{Hunter1986}; \citealp{Thronson1990}; \citealp{Hunter1993}}
and unusually large massive (OB) star population, including
a {\it galaxy-wide\/} surface density of Wolf-Rayet (WR) stars
$\ga$$4$ times that observed in the most active regions of star
formation in {\it any} other Local Group galaxy
\citep{Massey1998,Royer2001,Crowther2003}.
This ongoing star formation in IC~10 is expected to result in copious
X-ray emission, both from compact high-mass X-ray sources and
supernova-heated gas. The galaxy was observed twice using the {\it
ROSAT} HRI in 1996 (70.3~ks combined exposure), although the X-ray
flux in the HRI band is diminished by a factor of $\approx$5 due to
the Galactic column of $4.8\times 10^{21}$~cm$^{-2}$
\citep{Stark1992} and the expected absorption internal to IC~10 of
$\approx$2$\times 10^{21}$~cm$^{-2}$ \citep[e.g.,][]{Yang1993}. One
highly significant source was previously detected (hereafter X-1)
within the optical extent of IC~10 \citep{Brandt1997a}. This source
was found to vary by a factor of $\approx$3 on $\sim$1 day timescales,
with an average absorbed HRI flux of
$4\times10^{-13}$~erg~cm$^{-2}$~s$^{-1}$ (or an unabsorbed $L_{\rm
0.1-2.5~keV}\approx7\times 10^{37}$~erg~s$^{-1}$). The HRI-derived
position (5\arcsec\ rms) placed \hbox{X-1} in a region with intense star
formation as well as the most massive H~I cloud in IC~10. Notably, X-1
was found to lie $\approx$2\arcsec\ from [MAC92]~17 ($V=21.76$), an
emission-line source identified as a WR star by
\citet{Massey1992} and \citet{Crowther2003}, and within
$\approx$8\arcsec\ of the centroid of a non-thermal radio superbubble
\citep[$\approx$45\arcsec/150~pc diameter;][]{Yang1993}.  The
combination of the high X-ray luminosity and strong variability argued
that \hbox{X-1} is a powerful X-ray binary (containing a neutron star or
black hole).

In this Letter we report on {\it Chandra} and {\it HST}
follow-up observations of IC~10 that further characterize the
nature of X-1. In particular, these observations more accurately
determine the X-ray luminosity and spectrum of \hbox{X-1} (both of which were
uncertain since the {\it ROSAT} HRI had essentially no spectral
capability and was affected by the large absorption) and
confirm the likely association of \hbox{X-1} and the WR star [MAC92]~17
(there are several potential optical counterparts to \hbox{X-1} within the
5\arcsec\ radius HRI error circle). In $\S$\ref{reduction} we outline
the relevant X-ray and optical observations and analyses, while in
$\S$\ref{discussion} we summarize our findings and discuss \hbox{X-1} in the
broader context of black-hole binaries (BHBs).

\section{Observations and Analysis}\label{reduction}

\subsection{{\it Chandra} Observations}\label{xobs}

IC~10 was observed on 2003 March 12 with ACIS-S \citep[the
spectroscopic array of the Advanced CCD Imaging
Spectrometer;][]{Garmire2003}. To limit telemetry saturation in the
case of high background, only chips S2 and S3 in the ACIS-S array were
operated, while a 500-pixel subarray was used to lower the frame time
to 1.7~s and thus limit potential pileup of X-1. Events were
telemetered in Very Faint (VF) mode, and the CCD temperature was
$-120^{\circ}$~C. The optical extent of IC~10
($\approx$6\farcm3$\times$5\farcm1) is only marginally cropped by the
2-chip, 500-pixel subarray configuration (16\farcm8$\times$4\farcm1).

Analysis was performed using the CIAO software (v2.3) provided
by the {\it Chandra} X-ray Center (CXC), but also with FTOOLS
(v5.2) and custom IDL software. The data were corrected for the
radiation damage sustained by the CCDs during the first few months of
{\it Chandra} operations using the Charge Transfer Inefficiency (CTI)
correction procedure of
\citet{Townsley2002}.\footnote{For details see
http://www.astro.psu.edu/users/townsley/cti/.}  Following
CTI-correction, the data were reprocessed with the CIAO tool
ACIS\_PROCESS\_EVENTS to remove the standard 0\farcs5 pixel
randomization and to flag potential ACIS background events using VF
mode screening. We performed standard {\it ASCA} grade selection in
the 0.3--8.0~keV band, excluding all bad columns, bad pixels, VF-mode
background events, and cosmic-ray afterglows. We used only data taken
during times within the CXC-generated good-time intervals. The
background varied by $<$30\% during the observation, with the average
background rate being $2.2\times10^{-7}$~counts~s$^{-1}$~pixel$^{-1}$
and $7.0\times10^{-7}$~counts~s$^{-1}$~pixel$^{-1}$ on chips S2 and
S3, respectively (in agreement with ACIS quiescent-background
calibration measurements).\footnote{See
http://cxc.harvard.edu/contrib/maxim/bg/index.html.} The total net
exposure time is 29,191~s. 

X-ray sources were identified using the CIAO wavelet algorithm
WAVDETECT \citep{Freeman2002} on the \hbox{0.3--8.0~keV} (full),
\hbox{0.3--2.0~keV} (soft), and \hbox{2.0--8.0~keV} (hard) images with
a probability threshold of $10^{-6}$. In total, 43 sources were
detected. The positions, counts, and spectra for these sources were
derived using ACIS\_EXTRACT \citep[AE;][]{Broos2002} with 95\%
encircled-energy radii [based on point spread functions (PSFs) from
MKPSF and the PSF library]; the adopted X-ray source positions were
found using the matched-filter method within AE.

\subsection{{\it HST} Observations}\label{hstobs}

To compare the X-ray and optical emission, we observed IC~10 with the
{\it HST} Advanced Camera for Surveys \citep[ACS;][]{Ford1998} on 2002
October 12. IC~10 was imaged along its major axis
($\approx50^{\circ}$) using two ACS pointings separated by
130$\arcsec$. Each pointing consisted of $\approx0.5$-orbit exposures
in the $BVI$ bands (F435W, F555W, F814W). The standard pipeline data
products were used, although we mosaiced the two pointings ourselves
due to a problem with the pipeline MULTIDRIZZLE procedure. Optical
source positions and magnitudes were measured using SEXTRACTOR
\citep[v2.2.1;][]{Bertin1996}. 

\subsection{Astrometry of Optical and X-ray Images}\label{optobs}

Due to the small number of X-ray sources that overlap the ACS field of
view, we aligned both datasets to the set of high-quality ground-based
images available from the NOAO archive for IC~10 (PI:
Massey). Briefly, the NOAO images were matched to 55 GSC2.2 sources
within 10$\arcmin$ of IC~10 ($\Delta\alpha=0$\farcs20,
$\Delta\delta=-0$\farcs26), giving an rms scatter of 0\farcs19; this
alignment was confirmed using 11 Tycho~2 stars (0\farcs26 rms).  Given
the optical source density in the IC~10 region and apparent small,
systematic field distortions in the NOAO images, we only matched 332
{\it HST} sources brighter than $B=23$ within a 1\farcm5 region of X-1
($\Delta\alpha=5$\farcs32, $\Delta\delta=0$\farcs27). This gives an
astrometric accuracy of 0\farcs10 rms for the {\it HST} positions
within the vicinity of X-1. Finally, 14 of the 43 significant X-ray
sources (including X-1) have plausible optical counterparts, whereas
only two false matches are expected.
Aligning the X-ray and optical images ($\Delta\alpha=-0$\farcs44,
$\Delta\delta=0$\farcs53) yields an astrometric accuracy of 0\farcs28
rms for the X-ray positions.

\subsection{X-ray Analysis}\label{xanal}

\begin{figure}
\vspace{-0.0in}
\centerline{
\hglue0.2in\includegraphics[height=8.0cm, angle=-90]{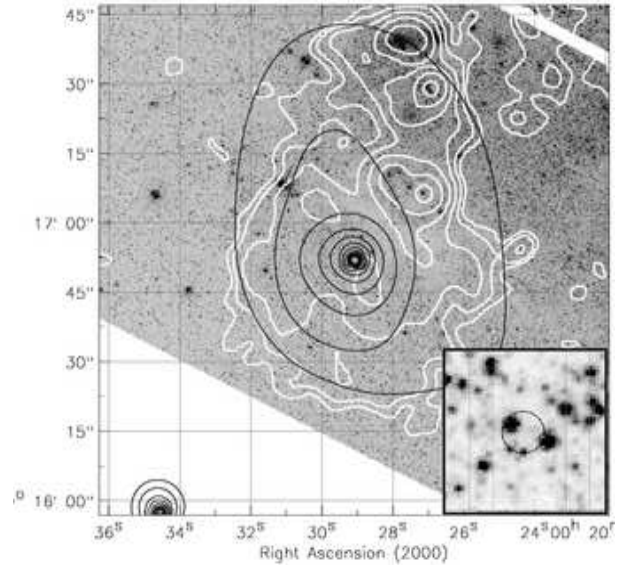}
}
\vspace{-1.2in} 
\figcaption{X-ray (dark) and radio (light) contours in the vicinity of 
X-1 overlaid on the {\it HST} ACS F814W image. The image shows \hbox{X-1} in
relation to the X-ray and radio emission from the surrounding
superbubble. {\it Inset:} 50-pixel ($\approx$2\farcs2) close-up of X-1
indicating the 0\farcs30 X-ray error circle (statistical$+$systematic)
relative to the ACS F814W counterparts. \hbox{X-1} lies 0\farcs23 from the
confirmed WR star [MAC92]~17A
\citep[slightly above and to the left of the X-ray
centroid;][]{Crowther2003}.
\label{fig:overlay}}
\vspace{-0.05in} 
\end{figure} 

IC~10~X-1 has 4403 counts in the full band and, based on our
astrometric solutions, it has a J2000 position of
$\alpha=$00$^{\mathrm h}$20$^{\mathrm m}$29\fs09,
$\delta=+$59\arcdeg16\arcmin51\farcs95. X-ray contours of \hbox{X-1}
are shown in Fig.~\ref{fig:overlay} overlaid on the {\it HST} ACS
F814W image. The Fig.~\ref{fig:overlay} inset indicates the {\it
Chandra} positional uncertainty of \hbox{X-1}, which is an order of
magnitude smaller than found previously with the {\it ROSAT}
HRI. \hbox{X-1} lies 0\farcs23 from J002029.1+591652.1 \citetext{an
$I_{AB}=21.8$ WR star identified as [MAC92]~17
by \citealp{Massey1992} and [MAC92]~17A by \citealp{Crowther2003}},
0\farcs33 from J002029.0+591651.6 ($I_{AB}=24.4$), 0\farcs37 from
J002029.1+591651.7 ($I_{AB}=24.0$), and 0\farcs42 from
J002029.0+591651.8 \citetext{an $I_{AB}=21.8$ OB supergiant identified
as [MAC92]~17B by \citealp{Clark2003}}. Given the astrometric uncertainty in
both the {\it HST} and {\it Chandra} coordinate frames (i.e.,
0\farcs30, much larger than the individual centroiding errors),
\hbox{X-1} is most likely associated with [MAC92]~17A. Although we
cannot exclude entirely an association with the latter sources (or
fainter undetected sources) based on the astrometric errors, the
probability of \hbox{X-1} lying within 0\farcs23 of a WR star by
chance is $\approx$0.05--0.2\%.\footnote{Narrow-band photometric
observations of IC~10 have identified $\sim100$ WR stars, of which 26
have been spectroscopically confirmed
\citep{Massey1995,Royer2001,Crowther2003}.}

\begin{figure}
\vspace{-0.1in}
\centerline{
\includegraphics[height=9.0cm]{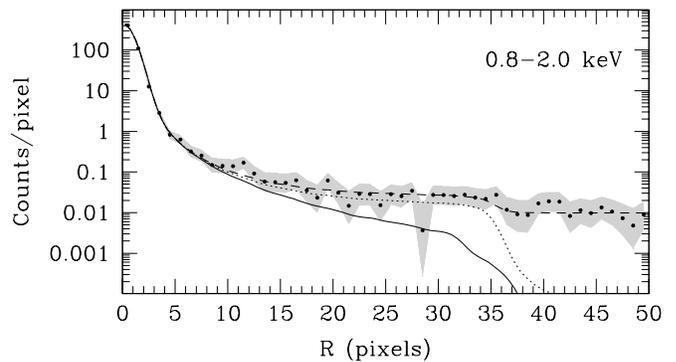}
}
\vspace{-3.7cm} 
\figcaption{The 0.8--2.0~keV ACIS-S radial profile (small filled
circles) of X-1, compared to our simple model with and without
background (dashed and dotted curves, respectively) and the {\it
Chandra} on-axis PSF (solid curve) calculated at 1.5~keV (i.e.,
closest to where the spectrum of \hbox{X-1} peaks). The shaded region
indicates the 1$\sigma$ deviation of the measured profile. Deviations
from the PSF can be seen out to radii of $\approx$35 pixels above
0.8--2.0~keV background of 0.010~counts~pixel$^{-1}$. A pixel is
equivalent to 0\farcs492.
\label{fig:X1-profile}}
\vspace{-0.15in}
\end{figure} 

In addition to an X-ray point source in Fig.~\ref{fig:overlay}, we
found faint extended X-ray emission roughly centered on \hbox{X-1} and
co-spatial with the \citet{Yang1993} non-thermal radio superbubble
(see Fig.~\ref{fig:overlay}). This X-ray emission extends out to
$\sim$20$\arcsec$ in the adaptively smoothed image. Given that the
X-ray emission appears azimuthally symmetric to $\sim$15$\arcsec$, we
extracted soft and hard-band counts around \hbox{X-1} in 1-pixel
annular bins. Due to the large foreground absorption column of IC~10,
we used the 0.8--2.0~keV band to reduce background where we detect no
signal. For comparison, we generated the on-axis ACIS-S PSF using
CHART,\footnote{See http://asc.harvard.edu/chart/.} extracting counts
in an identical manner to the data. The radial surface-brightness
distributions and 1$\sigma$ errors for the data are shown in
Fig.~\ref{fig:X1-profile} and demonstrate that \hbox{X-1} lies
systematically above the background out to radii of $\sim$35 pixels
\hbox{($\sim17\arcsec$)} and is clearly extended compared to the PSF
model. The hard-band profile is consistent with a point source and
background only.

We find 175 (129) counts in the full (soft) band between radii of
7--35 pixels (note that contamination from \hbox{X-1} should be $<1$\%
at such radii).
The fact that the extended emission is present almost exclusively in
the soft band further supports the reality of the extent. An emission
model consisting of both a point source and a uniform disk with a
$17\arcsec$ (60~pc) radius convolved with the PSF represents the
0.8--2.0~keV surface-brightness distribution well, although there is
clearly still some residual scatter due to possible clumping. The flux
ratio of the extended disk component to the point source is
$\approx2$\%. The relative softness of the extended emission is
consistent with a physical origin as supernova-heated hot gas (see
also the X-ray spectral analysis below).

\begin{figure}
\centerline{
\includegraphics[width=8.3cm,angle=0]{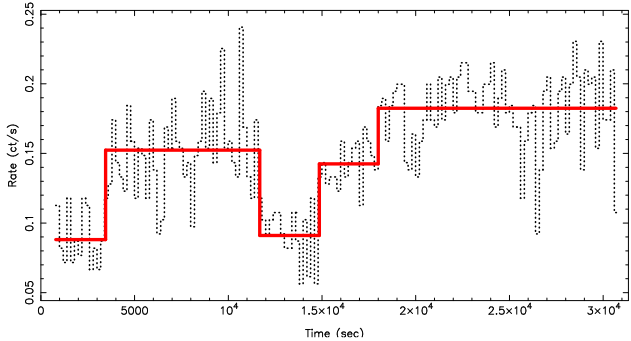}
}
\figcaption{X-ray light curve of IC~10~X-1 (dotted)
overlaid with the 99.9\% confidence level Bayesian ``blocks'' of
constant count rate (solid). The events are shown in 200~s bins.
\label{fig:X1-blocks}}
\vspace{-0.15in}
\end{figure} 

\citet{Brandt1997a} found that \hbox{X-1} varied by a factor of 
$\approx$3 during their {\it ROSAT} HRI observation but did not have
the statistics or the time sampling to constrain the variability
further. With $\approx$4400 counts over a contiguous 29.2~ks, our
light curve probes the short-term temporal properties better. A
Kolmogorov-Smirnov test indicates that \hbox{X-1} is variable at the
$>99.9$\% confidence level. To derive the amplitude and characteristic
timescale of variability, we used the Bayesian Block method provided
in the contributed CIAO software package SITAR.\footnote{See
http://space.mit.edu/CXC/analysis/SITAR/index.html}
Fig.~\ref{fig:X1-blocks} shows a binned light curve of \hbox{X-1},
overplotted with a series of contiguous "blocks" within which the
count rate is modeled as a constant at the 99.9\% confidence
level. \hbox{X-1} has peak, trough, and average count rates of 0.182
ct~s$^{-1}$, 0.088 ct~s$^{-1}$, and 0.151 ct~s$^{-1}$, respectively,
and exhibits rapid variability by as much as a factor of $\approx$2 in
$\la$3000~s. The hardness ratio, however, shows no significant
spectral variability. Since a significant fraction of X-ray binaries
are X-ray pulsars with pulse periods ranging from 0.01--1,000~s
\citep[e.g.,][]{White1995}, we generated power spectra to search for
pulsations using the robust $z^{2}_{1}$ method
\citep[e.g.,][]{Buccheri1983}. These power spectra also allowed us to
search for other periodic phenomena (e.g., eclipses or regular X-ray
bursts). No periodic behavior was found.

\begin{figure}
\centerline{
\includegraphics[height=7.0cm,angle=-90]{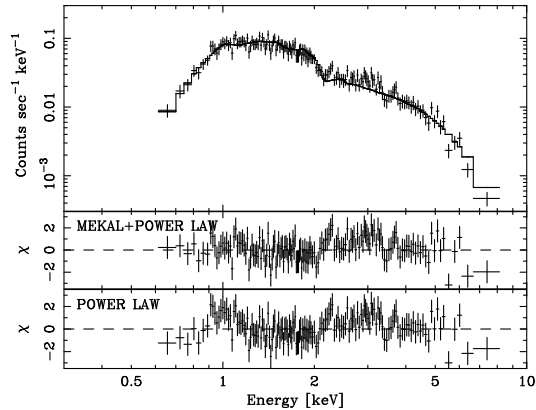}
}
\figcaption{
X-ray spectrum of the IC~10~X-1 point source, modeled with an
absorbed power law and thermal plasma with $\Gamma=1.83$,
$kT=1.49$~keV, and $N_{\rm H}=6.0\times10^{21}$. Note the large
residuals around 2.25~keV and 3.65~keV suggesting the presence of
possible emission lines.  
\label{fig:X1-spectra}}
\vspace{-0.05in}
\end{figure} 

The X-ray spectra were analyzed using XSPEC
\citep[v11.2;][]{Arnaud1996}. Unless stated otherwise, spectral
parameter errors are for the 90\% confidence level, assuming one
parameter of interest. The X-ray fluxes and absorption-corrected
luminosities for \hbox{X-1} were calculated from the spectral
fitting. For the detected \hbox{X-1} count rates above, pile-up is
estimated to be $\approx$15\%. Thus, spectral analysis of this source
must proceed with care. To model the spectrum in the presence of such
pile-up, we used the forward-modeling tool LYNX developed at PSU
\citep{Chartas2000}.  The ACIS spectrum of \hbox{X-1} was initially
fit with an absorbed power-law model, giving best-fit values of
$N_{\rm H}=(5.1\pm0.4)\times10^{21}$~cm$^{-2}$ and
$\Gamma=1.77\pm0.08$ ($\chi^{2}=97.6$ for 82 degrees of freedom). The
$\chi^{2}$ value and apparent systematic residuals (bottom panel in
Fig.~\ref{fig:X1-spectra}) indicate the model provides a poor fit to
the data. Given that we detect a low-level extended component around
\hbox{X-1}, we added an absorbed thermal plasma component \citep[{\it
mekal}; e.g.,][]{Mewe1985} to our spectral model. The resulting
best-fit values are $N_{\rm
H}=(6.0^{+0.20}_{-0.08})\times10^{21}$~cm$^{-2}$ (in agreement with
the expected Galactic$+$intrinsic absorption),
$\Gamma=1.83^{+0.07}_{-0.11}$, and $kT=1.49^{+0.15}_{-0.12}$~keV
($\chi^{2}=61.6$ for 80 degrees of freedom). The abundances were not
well constrained and were fixed at $Z=0.15Z_{\odot}$ (see
$\S$\ref{introduction}). While this model provides a significant
improvement to both the $\chi^{2}$ value ($>$99.99\% based on the
$F$-test) and the apparent systematic residuals below $\sim$2~keV,
there are still notable residuals around 2--4~keV (see middle panel in
Fig.~\ref{fig:X1-spectra}). These residuals, particularly around
2.25~keV and 3.65~keV, are perhaps due to marginally-resolved emission
lines from an \hbox{X-ray}-photoionized wind (e.g., as seen in
Cyg~X-3; see $\S$\ref{discussion}). The 0.5--8.0~keV absorbed flux and
unabsorbed luminosity of \hbox{X-1}, corrected for pileup, are
$1.57\times10^{-12}$ erg~cm$^{-2}$~s$^{-1}$ and $1.50\times10^{38}$
erg~s$^{-1}$, respectively. The thermal component contributes
$\approx$20\% of the total absorbed flux and $\approx$27\% of the
total unabsorbed luminosity, which is substantially more than was
estimated from fitting the radial profile. This suggests that either
much of this component arises from X-1 itself or that the extended
component is strongly centrally peaked rather than uniformly
distributed. We also analyzed the spectrum of the extended X-ray
component (3\farcs5--17\arcsec annulus around X-1). Given the small
number of counts in this region, we modeled its spectrum using the
Cash statistic \citep{Cash1979} instead of $\chi^2$. An absorbed {\it
mekal} model with fixed $N_{\rm H}=5.9\times10^{21}$~cm$^{-2}$ (as
found for X-1) gives abest-fit temperature of
$kT=0.87^{+0.17}_{-0.21}$~keV. The 0.5--8.0~keV absorbed flux and
unabsorbed luminosity of this extended component are
$1.73\times10^{-14}$ erg~cm$^{-2}$~s$^{-1}$ and $3.24\times10^{36}$
erg~s$^{-1}$, respectively. This flux is consistent with our
radial-profile fitting and supports an origin as hot gas associated
with the radio superbubble.

\section{Discussion}\label{discussion}

This {\it Chandra} observation of IC~10 has resulted in improvements
by factors of $\approx$15 in terms of spatial resolution and
astrometric accuracy and $\approx$10 in terms of statistics for
\hbox{X-1}, compared to previous {\it ROSAT} HRI exposures, allowing
substantially better constraints to be placed on the nature of this
object. The combination of the high luminosity and strong variability
clearly demonstrates that \hbox{X-1} is a powerful X-ray binary,
containing a black hole or neutron star. Its likely physical
association with [MAC92]~17A implies that the progenitor of \hbox{X-1}
must have evolved even more rapidly and been more massive than
[MAC92]~17A \citep[WNE stars typically have
$M\approx40$--50$M_{\odot}$ and $t_{\rm life}\la5$~Myr;
e.g.,][]{Maeder1994}. Thus, we expect \hbox{X-1} to be a BHB
\citep[although we note that there are clear counter examples; see,
e.g.,][]{Kaper1995}.  Optical spectroscopic monitoring of X-1 to
search for variability would be one of the best methods to prove the
WR-BHB interpretation; \citet{Clark2003} find hints of variability in
the $\lambda$4686 feature of [MAC92]~17A, but longer dedicated
observations are needed to secure this finding.

The X-ray spectrum of X-1 is acceptably fit with a $\Gamma=1.83$ power
law and $kT=1.49$~keV thermal plasma \citep[potentially indicative a
BHB in a hard state; e.g.,][]{McClintock2003}. The residuals of X-1 in
the 2--4~keV range hint at further spectral complexity, perhaps from
emission lines or absorption features.  Longer {\it Chandra} or {\it
XMM-Newton} observations of X-1 would also be useful to quantify the
origin of the spectral residuals, as well as to place additional
constraints on the nature of the X-ray emission (e.g., to rule out
further the presence of pulsations or bursts).

The properties of \hbox{X-1} bear several similarities to the Galactic
X-ray binary Cyg~X-3 in terms X-ray luminosity, spectrum, and
variability \citep[e.g.,][]{Kitamoto1994, Liedahl1996, Predehl2000} as
well as donor type \citep[WNE vs. WN7;][]{vanKerkwijk1996,
Crowther2003}. Moreover, Cyg~X-3 has a complex X-ray spectrum
including numerous emission lines and a 9~keV edge that are attributed
to an X-ray-photoionized wind; the residuals seen from X-1 may have a
similar origin. Thus IC~10~X-1 is the only other confirmed example
of the short-lived WR X-ray binary.

The discovery of soft extended emission co-spatial with the large
non-thermal radio superbubble surrounding X-1 is important for
determining the nature of the bubble, although the poor statistics
obtained here do not allow strong constraints. Obvious comparisons can
be made with 30 Doradus \citep{Dennerl2001}, as well as with SS433/W50
\citep{Safi-Harb1997} and IC~342~X-1 \citep{Roberts2003}, all of which
have similar physical extents and radio/X-ray luminosities as the
bubble in IC~10. While the radio and X-ray emission from the IC~10
bubble is consistent with multiple supernovae, the superbubble does
lie off the $\Sigma$--$d$ relation \citep{Yang1993}, suggesting that
something else (perhaps X-1) may be powering the expansion. For
instance, in the case of W50, SS~433 is believed to contribute to the
expansion. High-resolution radio imaging of IC~10 could be used to
search for any jets associated with X-1 and explore whether X-1 has
had any influence on the nonthermal superbubble.

\acknowledgements 

We thank L.~Townsley and M.~van~Kerkwijk for several helpful
discussions, G.~Chartas for help with the forward-modeling of the
X-ray spectrum, M.~Muno for help with variability constraints,
S.~Clark and P.~Crowther both for interesting discussions regarding
X-1 and communicating their results prior to publication. We
gratefully acknowledge the financial support of CXC grant GO3-4112X
(FEB, WNB), STScI grant HST-GO-09683.01A (FEB, WNB) and NASA LTSA
grant NAG5-13035 (WNB).

\end{document}